\documentclass[twocolumn,showpacs,prb,footinbib,a4paper,superscriptaddress,floatfix]{revtex4}

\usepackage{amssymb,amsmath}
\usepackage{graphics}
\usepackage{dcolumn}
\usepackage{bm}
\usepackage{epsfig}

\begin{document}

\title{Nonequilibrium density matrix for quantum transport:\\ Hershfield approach as a 
McLennan-Zubarev form of the statistical operator}

\author{H. Ness}
\email{herve.ness@york.ac.uk}
\affiliation{Department of Physics, University of York, Heslington, York YO10 5DD,UK}
\altaffiliation{European Theoretical Spectroscopy Facility (ETSF), www.etsf.eu}

\begin{abstract}
In this paper, we formally demonstrate that the non-equilibrium
density matrix developed by Hershfield for the steady state has the form of a
McLennan-Zubarev non-equilibrium ensemble.
The correction term in this pseudo equilibrium Gibbs-like ensemble is directly
related to the entropy production in the quantum open system.
The fact the both methods state that a non-equilibrium steady state can be
mapped onto a pseudo-equilibrium, permits us to develop non-equilibrium quantities
from formal expressions equivalent to the equilibrium case. We provide an example:
the derivation of a non-equilibrium distribution function for the electron population 
in a scattering region in the context of quantum transport. 
\end{abstract}


\maketitle

\section{Introduction}
\label{sec:intro}

The understanding of irreversible phenomena including non-equilibrium (NE) steady
state is a longstanding problem of statistical mechanics.
The task of NE statistical mechanics is to understand
and describe how a system, initially at thermodynamical equilibrium, will try 
to respond and adjust to an external stimulus by evolving towards a new macroscopic
state that is compatible with this external constraint.
This involves the understanding of the transient and steady state regimes, as well
as the derivation of the corresponding kinetic and balance equations of NE thermodynamics.
  
Since Gibbs formulation of the method of statistical ensembles for equilibrium
many-body systems, it has been expected that some formal advantages may be
given by an approach to NE processes in which the Gibbs
ensembles play a prominent role \cite{McLennan:1959}.
The construction of such Gibbs-like ensembles for the NE steady state has been
explored by many authors.

On one hand, early attempts have been performed by McLennan \cite{McLennan:1959} for 
classical systems and by Zubarev 
\cite{Zubarev:1974,Zubarev:1994,Zubarev:1996,Zubarev:1997,Morozov:1998}
for both classic and quantum systems.
In simple terms, it is found that the steady state ensembles can be expressed
in terms of the external forces which maintain the deviation from equilibrium.
In Zubarev's formulation of NE steady state, the Gibbsian statistical mechanics
method is extended to include steady-state boundary conditions in the density-matrix
leading to his so-called NE statistical operator (SO) method (NESOM).
Such method consists in constructing a time-independent density matrix (statistical
operator) by solving an equation of motion with the proper NE boundary conditions.

A rigorous analysis of the existence and stability of such NE steady state, i.e.
its independence on the way the division into subsystems and reservoirs is performed
and its stability against local perturbations, have been performed using
$C^*$ algebraic methods in Refs.~[\onlinecite{Ruelle:2000,Tasaki:2003,Tasaki:2006,Tasaki:2011}].
Furthermore, rigorous definition of MacLennan-Zubarev ensembles have been 
given in Refs.~[\onlinecite{Tasaki:2003,Maes:2010}].

There is also an extensive literature which shows that the NESOM turns out to be very
convenient for concrete application (for example see the review Ref.~[\onlinecite{Morozov:1998}]
and the references therein).
More recently, applications of the McLennan-Zubarev form of the NE density matrix
have been done in the context of quantum electron transport.
For example, the problem of quantum transport for non-interacting electrons in effective
one-dimensional systems can be found in Ref.~[\onlinecite{Bokes:2003,Bokes:2005}], where the authors
rederive Zubarev's NESOM from a maximum entropy principle, since by essence a Gibbs
state is characterized by the principle of maximum entropy at fixed energy 
(see also Ref.~[\onlinecite{Zubarev:1994}]).
Applications for interacting electron-nuclei systems is provided in Ref.~[\onlinecite{Wang_Y:2007}],
in which the authors derive the kinetic equations governing time evolution of positions and 
momenta of atoms (in the classic limit) interacting with a quantum electron gas using the NESOM.

On the other hand, in the early nineties, Hershfield reformulated the problem
of NE steady state quantum statistical mechanics (QSM) in Ref.~[\onlinecite{Hershfield:1993}].
This was done by rewriting the conventional 
perturbation theory of NE QSM for the steady state regime in a form similar to that of 
an equilibrium QSM (called below a pseudo equilibrium).
In this reformulation, an explicit expression for the NE density matrix was provided as
well as a scheme upon which one can build non-perturbative calculations in
NE quantum systems.
This approach has permitted us to understand more clearly the non-equilibrium ensembles, 
and how NE boundary conditions can be imposed as a statistical operator. 
It has also been successfully applied in numerical applications
for the problem of quantum transport, in the presence or absence of interaction
between electrons \cite{Schiller:1995,Schiller:1998}.
Other applications have been performed by Han and co-workers 
in the context of quantum transport for electron-phonon interaction in quantum dots \cite{Han:2006},
for strongly correlated electron systems within a slave boson approach \cite{Han:2007}.
Han and co-workers also developed an equivalent formulation within the framework of 
the imaginary time formalism \cite{Han:2007b,Han:2010,Han:2010b,Dutt:2011,Han:2012}.

In this paper, we show and prove that Hershfield approach for the NE density matrix
is actually a specific form of the McLennan-Zubarev NESOM. 
By specific, we mean
that Hershfield approach can be seen as a particular case of the NESOM applied to
the problem of quantum transport, where a central scattering region (with
interaction or not) is connected to two (or more) leads and the whole system is 
at the same temperature $T$.
Our work provides a clear and formal connection between these two approaches 
which are widely used for applications in quantum transport.

Both McLennan-Zubarev and Hershfield show that the properties of a NE steady state
can be obtained in a formally equivalent manner as in an equilibrium state but
using a NE density matrix in a Gibbs form instead of the equilibrium 
Gibbs statistical ensembles. 
Hence it is also possible to derive NE quantities, such as distribution functions, 
from formal expressions used in the equilibrium case.
We consider the development of such distribution functions for the population 
of electrons in the context of quantum transport in the last part of the paper.

The paper is organised as follow.
In Sec.~\ref{sec:NESOM} and Sec.~\ref{sec:Hershfield}, we briefly recall the main
ingredients of the McLennan-Zubarev NESOM and of Hershfield approach respectively.
Sec.~\ref{sec:connection} is the main part of the paper where we formally establish
the connection between the two methods.
In Sec.~\ref{sec:NEdistrib} we derived the expression for the NE steady state distribution
of the electron population  of a central region (in the presence of interaction) connected 
to two (non-interacting) electron reservoirs.
Finally, we discuss further developments and present our conclusion in Sec.~\ref{sec:ccl}.

\section{McLennan-Zubarev non-equilibrium statistical operator method}
\label{sec:NESOM}

Classic and quantum statistical mechanics should provide microscopic foundations for the thermodynamics
description of many-body systems. For the equilibrium case, the method of statistical ensembles
developed by Gibbs gives a rigorous formulation of the thermodynamic quantities and relations.
Within the same line of reasoning, an extension of Gibbs method to the non-equilibrium cases would
permit to formulate the basic postulates of irreversible thermodynamics.
Such a formulation of non-equilibrium statistical mechanics has been provided by McLennan \cite{McLennan:1959}
and Zubarev \cite{Zubarev:1974,Zubarev:1994,Zubarev:1996,Zubarev:1997,Morozov:1998}.

For a system composed of $N$ independent parts (with $j=1,...,N$ Hamiltonian $H_j$, 
at temperature $\beta_j$, and with $\lambda=1,...,L$ species of particle $\lambda$ with 
number $N_j^{(\lambda)}$ and chemical potential $\mu_j^{(\lambda)}$) which are interacting 
by an interaction $W$,
the McLennan-Zubarev form of the NE statistical operator is given by
\begin{widetext}
\begin{equation}
\label{eq:rhoNESOM}
\rho = \frac{1}{Z}
{\rm exp} 
\left\{
-\sum_{j=1}^N \beta_j 
\left[
H_j - \sum_{\lambda=1}^L \mu_j^{(\lambda)} N_j^{(\lambda)}  \right]
- \int_{-\infty}^0 ds e^{\eta s} J_S(s)
\right\} \ ,
\end{equation}
\end{widetext}
where $Z$ is the normalisation factor $Z={\rm Tr}[\rho]$ and
the quantity $J_S(s)$ is obtained from $J_S(s) = \sum_j \beta_j J^q_j(s)$
with
$J^q_j(s)$ being the so-called non-systematic energy flow \cite{Tasaki:2006}, 
or heat flow, to the $j$th subsystem defined as
\begin{equation}
\label{eq:heatflow_NESOM}
J^q_j(s) =  \frac{d}{ds} \left( H_j(s) - \sum_\lambda \mu_j^{(\lambda)} N_j^{(\lambda)}(s) \right) \ .
\end{equation}

The operators are given in the Heisenberg representation, with the total
Hamiltonian $H=\sum_j H_j +W$ and
$H_j(s) = e^{iHs} H_j e^{-iHs}$, $N_j(s) = e^{iHs} N_j e^{-iHs}$.
A convergence factor $e^{\eta s}$ ($\eta > 0$) is introduced in the time
integral, where the limit $\eta\rightarrow 0$ is taken in the end, after all the calculations
are done.

The quantity $J_S(s)$ being the sum of heat flows divided by
subsystem temperatures, is therefore the entropy production rate of the whole system
\cite{Zubarev:1994,Tasaki:2006}.

\section{Hershfield approach for non-equilibrium density matrix}
\label{sec:Hershfield}

Hershfield reformulated the problem of NE steady state in quantum
statistical mechanics \cite{Hershfield:1993}
by developing an iterative scheme for the NE density matrix expressed in
terms of a series of power of $(W)^n$ where $W$ is the perturbative
part of the total Hamiltonian $H=H_0 + W e^{\eta t}$
that drives the system out of equilibrium
(and eventually also contains the interaction between the particles).

The expectation value of any operator $A$ in a NE steady state is then obtained
from a pseudo equilibrium as follows:
\begin{equation}
\label{eq:aveA}
\langle A \rangle = \frac{1}{Z^{\rm NE}} {\rm Tr}[\rho^{\rm NE} A] \ ,
\end{equation}
with the NE density matrix
\begin{equation}
\label{eq:rhoNE}
\rho^{\rm NE} = e^{-\beta (H-Y)} ,
\end{equation}
and the partition function $Z^{\rm NE}={\rm Tr}[\rho^{\rm NE}]$.

In the interaction representation (where the operators $X$ are given by 
$X_I(t)=e^{i H_0 t} X e^{-i H_0 t}$), the density matrix follows the usual
equation of motion
\begin{equation}
\label{eq:denmat}
\frac{\partial \rho_I(t)}{\partial t} = i [ \rho_I(t), W_{I}(t) ] \ .
\end{equation}
Hershfield introduced a new set of operators $Y_n$ which are of the order $\mathcal{O}(W^n)$
and from which the density matrix can be constructed by an iterative scheme.

The individual operators $Y_n$ follows the same differential equation as the density matrix
but in a recursive way:
\begin{equation}
\label{eq:Yn}
\frac{\partial Y_{n+1,I}(t)}{\partial t} = i [ Y_{n,I}(t), W_{I}(t) ] \ .
\end{equation}
The index of the operators $Y_n$ differs in each side of Eq.~(\ref{eq:Yn}) in order to
have the same power of the perturbation $W$ on both sides.

The differential equation Eq.~(\ref{eq:Yn}) can also be rewritten in terms of commutators as
\begin{equation}
\label{eq:Yn_bis}
[ H_0 , Y_n ]  - i \eta Y_n = [ Y_{n-1}, W ] \ ,
\end{equation}
where the positive infinitesimal $\eta$ is included to make the equation well defined \cite{Hershfield:1993}.

The operator $Y$ is then obtained from the sum $Y=\sum_{n=0}^\infty Y_n$.
The initial expression of the $Y_n$ operators is given by
$Y_0=\sum_i \mu_i N_i$. The important difference between the equilibrium and NE cases is that
the operator $Y_0$ does not commute with the perturbation $W$.
Furthermore, Hershfield showed that the full operator $Y$ and the total Hamiltonian $H$
commute in the limit of adiabatic switching of the perturbation ($\eta\rightarrow 0^+$). 
Because $Y$ and $H$ commute, Hershfield interpreted the $Y$ operator as the operator into 
which $Y_0$ ``evolves'' under the action of the perturbation $W$.

We show in the next section that the NE density matrix $\rho^{\rm NE}$ with the presence of the $Y$ operator
is actually a McLennan-Zubarev form of a NE statistical operator 
(for a system at the same temperature $kT=1/\beta$).

\section{Hershfield density matrix as a McLennan-Zubarev form of the statistical operator}
\label{sec:connection}

We now rewrite the McLennan-Zubarev NE statistical operator for the conditions considered by Hershfiedl,
i.e. one specie of particle (electrons) $L=1$ and a unique temperature $\beta_j=\beta$. Hence
Eq.~(\ref{eq:rhoNESOM}) becomes
\begin{widetext}
\begin{equation}
\label{eq:rhoNESOM_H}
\rho = \frac{1}{Z}
{\rm exp} 
\left\{- \beta \sum_{j=1}^N \left( H_j - \mu_j N_j  
- \int_{-\infty}^0 ds e^{\eta s} J^q_j(s) \right)
\right\} 
= 
\frac{1}{Z} e^{-\beta (H-\Upsilon)}
\ .
\end{equation}
\end{widetext}
In the second equality of Eq.~(\ref{eq:rhoNESOM_H}), we have rewritten the NE statistical operator 
in the form of a NE density matrix with
\begin{equation}
\label{eq:Upsilon}
\Upsilon = Y_0 + W + \int_{-\infty}^0 dx e^{\eta x} e^{iHx} i [W, H_0 - Y_0] e^{-iHx} \ ,
\end{equation}
with
the total Hamiltonian $H=\sum_j H_j +W=H_0+W$, 
and $Y_0=\sum_j \mu_j N_j$,
and 
\begin{equation}
\label{eq:J_S}
\begin{split}
\sum_j \frac{1}{\beta} J^q_j(x)
                        & =  \sum_j \frac{d}{dx} \left( H_j - \mu_j N_j \right)(x) \\
		 	& =  \frac{d}{dx} \left( H_0(x) - Y_0(x) \right) \\ 
                        & = i [H, H_0(x) - Y_0(x)] \\
                        & = e^{iHx} i [W, H_0 - Y_0] e^{-iHx} \\
                        & = i [W(x), H_0(x) - Y_0(x)] \ .
\end{split}
\end{equation}
We use the fact that the operator $Y_0$ commutes with the unperturbed non-interacting
Hamiltonian $H_0$.

To prove that the Hershfield approach is actually a McLennan-Zubarev form of the NE statistical operator,
we have to prove that the operator $\Upsilon$ is just the operator $Y$ in Hershfield method.

For that, we expand the time dependence of the commutator $A=i [W, H_0 - Y_0]$ in a series expansion
\begin{equation}
\label{eq:Ax}
\begin{split}
A(x) & =  e^{iHx} A e^{-iHx} \\
     &  = A + [iHx,A] + \frac{1}{2} [iHx,[iHx,A]] \\ 
     & + \frac{1}{3} [iHx,[iHx,[iHx,A]]] + ... 
\end{split}
\end{equation}
of powers of $(W)^n$, knowing that $H=H_0+W=\mathcal{O}(W^1)$ and $A=\mathcal{O}(W^1)$.

It is then natural to expand, as in Hershfield approach, the operator $\Upsilon$ in a series
$\Upsilon=\sum_n \Upsilon_n$ where each term  $\Upsilon_n$ corresponds to a power $W^n$.
The aim of the derivation is to identify the terms of each order of the perturbation $W$ in
the interaction representation scheme of Hershfield for $Y_{n,I}(t)$ and in the Heisenberg
representation used for the expression of $\Upsilon$ in the NESOM. This is easily
done for the lowest order terms.

At the zero-th order of the perturbation, it is clear from Eq.~(\ref{eq:Upsilon}) that
$ \Upsilon_0 = Y_0$. For the higher order, it is convenient to generalise Eq.~(\ref{eq:Upsilon})
as 
\begin{equation}
\label{eq:Upsilon_t}
\Upsilon(\tau) = Y_0 + W + \int_{-\infty}^\tau dx e^{\eta s} e^{iHx} i [W, H_0 - Y_0] e^{-iHx} \ ,
\end{equation}
and take the limit $\tau=0$ in the end to make the connection between the NESOM and Hershfield
approach.
Hence we have
\begin{equation}
\label{eq:Upsilon_tder}
\frac{\partial\Upsilon(\tau)}{\partial\tau} 
=  e^{iH\tau} i [We^{\eta\tau} , H_0 - Y_0] e^{-iH\tau} + \frac{\partial Y_0}{\partial\tau} 
   + \frac{\partial W}{\partial\tau} \ ,
\end{equation}
with $\partial_\tau Y_0 = i [H,Y_0(\tau)]$ and $\partial_\tau W = i [H,W(\tau)]$.

To get the term linear in $W$, we have to consider the lowest order expansion
in Eq.~(\ref{eq:Ax}) for the time
evolution operator in terms of the non-interacting Hamiltonian $H_0$ only, i.e.
$H \rightarrow H_0$ and $X(\tau) \rightarrow X_I(\tau)$.
Hence the right hand side of Eq.~(\ref{eq:Upsilon_t}) becomes
$i [W_I(\tau) , H_0 - Y_{0,I}(\tau)] + i [H_0,W_I(\tau)]$, where the term $e^{\eta\tau}$ is
included in $W_I(\tau)$.
For the left hand side of Eq.~(\ref{eq:Upsilon_t}), we assume that the Heisenberg representation
of $\Upsilon$ can rearrange as $e^{iH_0\tau} [$ sum of terms in $\mathcal{O}(W^n)] e^{-iH_0\tau}$
i.e. $e^{iH_0\tau} [\sum_n \Upsilon_n] e^{-iH_0\tau}$.
Hence at the lowest order in $W$, we get the interaction representation of $\Upsilon_{n=1}$, 
and therefore we find the lowest order version of Eq.~(\ref{eq:Yn}) for  $\Upsilon_n$:
\begin{equation}
\label{eq:Upsilon_n1}
\partial_\tau \Upsilon_{1,I}(t) = - i [ W_{I}(t) , Y_{0,I}(\tau) ] \ .
\end{equation}

The same result can also be obtained more directly from Eq.~(\ref{eq:Upsilon}) by considering
the lowest order expansion of the time evolution operator:
\begin{equation}
\label{eq:Upsilon_n1_bis}
\Upsilon_1 = W + \int_{-\infty}^0 dx e^{\eta s} e^{iH_0x} i [W, H_0 - Y_0] e^{-iH_0x} \ ,
\end{equation}
and integrating by part the term in $e^{iH_0x} i [W, H_0 ] e^{-iH_0x} = - \partial_x W_I(x)$
to find
\begin{equation}
\label{eq:Upsilon_n1_ter}
\Upsilon_1 = i \int dx e^{iH_0x} [Y_0, W ] e^{-iH_0x}\ ,
\end{equation}
which is just the integrated version of Eq.~(\ref{eq:Upsilon_n1}) or Eq.~(\ref{eq:Yn}).

The higher order terms $\Upsilon_{n \ge2}$ can be found from
\begin{equation}
\label{eq:Upsilon_nge2}
\Upsilon_{n \ge 2} = i \int_{-\infty}^0 dx e^{iHx} [We^{\eta s} , H_0 - Y_0] e^{-iHx} \ ,
\end{equation}
however the derivation is much more cumbersome that for the lowest order terms.

Instead, one can use Eq.~(\ref{eq:Upsilon}) and perform an analysis and decomposition order
by order of the powers in $\mathcal{O}(W^n)$.
For that we first rewrite Eq.~(\ref{eq:Upsilon}) as
\begin{equation}
\label{eq:Upsilon_A}
\begin{split}
\Upsilon = Y_0 + W -  \int_{-\infty}^0 dx \partial_x W(x)  \\
- i \int_{-\infty}^0 dx e^{iHx} [We^{\eta x} , Y_0] e^{-iHx} \ ,
\end{split}
\end{equation}
using the fact that 
$e^{iHx} i [W, H_0] e^{-iHx}
=e^{iHx} i [W, H] e^{-iHx} = i [W(x), H] = - \partial_x W(x)$. 
Note that when not explicitly written, the term $e^{\eta x}$ is included in the
perturbation $W$. Finally, using the fact that, in leading order, the operator $Y$ in 
Hershfield approach is the time evolution of $Y_0$, $Y=e^{iHx} Y_0 e^{-iHx}$, we get
in leading order
\begin{equation}
\label{eq:Upsilon_B}
\Upsilon = Y_0 - i \int_{-\infty}^0 dx [W(x) , Y(x)]  \ .
\end{equation}
Hence again, in leading order, we find that by expanding Eq.~(\ref{eq:Upsilon_B}) in
powers of $\mathcal{O}(W^n)$, we keep only the time dependence in terms of $H_0$ in the
series expansion, and we find $\Upsilon_0 = Y_0$, 
$\Upsilon_1 = - i \int_{-\infty}^0 dx  [W_I(x) , Y_{0,I}(x)]$, and
$\Upsilon_{n+1} = - i \int_{-\infty}^0 dx [W_I(x) , Y_{n,I}(x)]$ which is the integrated
expression of the right hand side of Eq.~(\ref{eq:Yn}).

Finally to conclude this section, we consider Eq.~(\ref{eq:Upsilon_tder}) and rewrite it
in terms of commutators to find that:
\begin{equation}
\label{eq:Upsilon_C}
\begin{split}
 & \partial_\tau\Upsilon(\tau)= \\
 & i [W(\tau) , H_0(\tau) - Y_0(\tau)] + i [H,Y_0(\tau)]+ i [H,W(\tau)] \\
= & i [W(\tau) , - Y_0(\tau)] + i [H,Y_0(\tau)] \\
= & i [H_0(\tau),Y_0(\tau)]
=  e^{iH\tau} i [H_0,Y_0] e^{-iH\tau} = 0 \ .
\end{split}
\end{equation}
Hence $\Upsilon(\tau)=\Upsilon$ is constant of motion, and 
$\partial_\tau\Upsilon(\tau)=i[H,\Upsilon(\tau)]=0$ implies that the operator $\Upsilon$
commutes with the total Hamiltonian $H$, as the Hershfield operator $Y$ commutes with $H$.

Therefore we have shown that the NE density matrix $e^{-\beta(H-Y)}$ of Hershfield approach 
is indeed a McLennan-Zubarev form of the NE statistical operator.
Finally, we can note that the NE density matrix/statistical operator depends, via the 
operator $Y/\Upsilon$, on the NE conditions as expected, i.e. on the different chemical
potentials $\mu_i$ in $Y_0$, but also on the interaction $W$ and on how the initial
$Y_0$ evolves under the perturbation $W$.

\section{An application for non-equilibrium distribution}
\label{sec:NEdistrib}

The fact that the NE steady state can be described as a pseudo equilibrium state, 
with a modified Gibbs-like statistics, permit us to determine 
NE quantities from formal expressions used in the equilibrium case
(compare Eq.~(\ref{eq:GF_feq}) and Eq.~(\ref{eq:GF_fNE}) below).
In this section, we derive an expression for the NE distribution function 
of the electron population in a central region connected to two reservoirs.
The study of other NE thermodynamical quantities for non-interacting quantum transport
(current-induced forces and thermodynamical potentials) has been addressed in
[\onlinecite{Todorov:2000,Todorov:2004,DiVentra:2004,Hyldgaard:2012}]. 
 
As an example, we consider in the following the NE distribution function of a central 
region consisting of a single level (with interaction) 
connected to two (left and right) reservoirs at their own equilibrium.
The statistics in each reversoir is given by the Fermi-Dirac distribution with 
chemical potentials $\mu_L$ and $\mu_R$ and temperature $T_L$ and $T_R$. 
In the NE conditions $\mu_L \ne \mu_R$ and/or $T_R \ne T_R$.

The Hamiltonian for the central scattering region $C$ is simply given by 
$  H_C   = \varepsilon_0 d^\dagger d $
where $d^\dagger$ ($d$) creates (annihilates) an
electron in the level $\varepsilon_0$.
The specific model used for the leads connected to the central region
does not need to be specified, as long as the leads can be described by 
an embedding self-energy $\Sigma_{\alpha}$ in the electron GF 
of the central region ($\alpha=L,R$).

At equilibrium, the average of the electron population of a single level
coupled to a thermal bath  $\langle d^\dagger d \rangle$ leads to the
equilbrium Fermi-Dirac distribution $f^{\rm eq}$.
For NE conditions, the average $\langle d^\dagger d \rangle$ as given
by Eq.~(\ref{eq:aveA}) is difficult to derive exactly especially in the
presence of interaction \cite{Schiller:1995,Schiller:1998,Han:2006}.
However, because of the pseudo equilibrium nature of the NE steady
state statistics,
we can assume that such an average is well behaved and leads to a NE 
distribution $f^{\rm NE}$.

In order to obtain a compact form for $f^{\rm NE}$, we have found that,
instead of calculating the series expansion of the operator $Y$, a more straight 
forward approach is obtained by using NE Green's functions (GF) in the 
steady state regime \cite{Note:1}.
The GF are correlation functions whose thermodynamical averages are formally 
identical to those calculated in Hershfield approach. 

Both perturbation series used in the NE GF approach and in the derivations 
of the equations for the Y operator in Hershfield approach start from 
the same nonequilibrium series expansion. They are just two different ways 
of summing that series. 
For a noninteracting problem for which the series can be resumed
exactly, the NE GF and the Hershfield Y operator approach provide the same 
result \cite{Schiller:1995,Schiller:1998}. 
For an interacting system, one must resort to approximations to partially
resum the series, and therefore the two approaches are similar only when the
same approximations are used.

The different GF in the central region can be obtained from two correlation functions 
(i.e. the so-called lesser and greater GF):
\begin{equation}
\label{eq:GF}
\begin{split}
G^<(t,t') & = -i \langle d^\dag(t') d(t) \rangle \ , \\ 
G^>(t,t') & =  i \langle d(t) d^\dag(t') \rangle \ , 
\end{split}
\end{equation}
where $d^\dag$ ($d$) creates (annihilates) an electron in the single level of the central region and
$\langle\dots\rangle$ is the average over the proper equilibrium or NE ensemble, as given
in Eq.~(\ref{eq:aveA}).

The other GF, the advanced and retarded GF, are obtained from the combination
of the lesser and greater components as
\begin{equation}
\label{eq:GFra}
G^{r/a}(t,t')=\pm\theta(\pm(t-t'))(G^>(t,t')-G^<(t,t')) \ .
\end{equation}

The interaction in the central region is obtained from a perturbation expansion,
via partial resummation of Feynmann diagrams, and enters the definition of the GF 
via the self-energy $\Sigma_{\rm int}$ in the
Dyson equations of $G^{r,a}$ and in the quantum kinetic equations of $G^\lessgtr$.

At equilibrium and in the steady state, all quantities depend only on the time difference
$X(t,t')=X(t-t')$ and can be Fourier transformed in an single-energy representation $X(\omega)$.

At equilibrium, from the relation 
$G^>-G^<=G^r-G^a$ and the KMS condition
\cite{Kubo:1957,Kubo:1966,Abrikosov:1963,Fetter:1971,Bratteli:1997} 
$G^>(\omega)= - e^{\beta(\omega-\mu^{\rm eq})} G^<(\omega)$, one
recovers the conventional relation 
\begin{equation}
\label{eq:GF_feq}
G^< = - f^{\rm eq} (G^> - G^<) = - f^{\rm eq} (G^r - G^a) \ ,
\end{equation}
with $f^{\rm eq}(\omega)=[1-G^>/G^<]^{-1}=[1+e^{\beta(\omega-\mu^{\rm eq})}]^{-1}$ being the
equilibrium Fermi-Dirac distribution.
The equilibrium KMS condition arises from the fact that the statistical operator $e^{-\beta H}$
looks formally like the time evolution operator
$e^{-iHt}$ if one works with imaginary time $t \equiv -i \beta$. 
In general, for any two operators $A$ and $B$, the KMS relation is given by
$\langle A(t-i\beta) B(t') \rangle = \langle B(t') A(t) \rangle$
\cite{Kubo:1957,Kubo:1966,Abrikosov:1963,Fetter:1971,Bratteli:1997}.

In the NE steady state the situation is different. Even if the steady state can be
seen as a pseudo equilibrium state, with a statistical operator $e^{-\beta(H-Y)}$,
the KMS relation is modified as follows
$\langle A(t-i\beta) B(t') \rangle = \langle e^{-\beta Y } B(t') e^{\beta Y} A(t) \rangle$.
Depending on the nature of the operator $B$, additional contributions arise from the expansion
$ e^{-\beta Y } B e^{\beta Y} = B + [-\beta Y, B] + [-\beta Y, [-\beta Y, B] ] / 2! + \dots$.

However, because of the intrinsic pseudo equilibrium nature of the NE steady state,
it is entirely justified to use the Gibbs-like ensemble, provided by either Hershfield
or McLennan-Zubarev methods, to define the NE distribution $f^{\rm NE}$ for the relationship 
between the GFs in a similar way as done for the equilibrium relation.
That is, we can extend the formal definition of the equilibrium distribution 
to the NE conditions, i.e. the distribution $f^{\rm NE}$ of 
the electron population in the 
NE steady state \cite{Ness:2009,Ness:2010,Ness:2012} :
\begin{equation}
\label{eq:GF_fNE}
G^<(\omega)  = -  f^{\rm NE}(\omega)   \left( G^r(\omega) - G^a(\omega) \right) \ ,
\end{equation}
where we are now considering full NE GF \cite{Note:2}. This is a rigorous definition
for the NE steady state, and not an ansatz.

With respect to Refs.~[\onlinecite{Todorov:2000,Todorov:2004,Hyldgaard:2012}], 
the NE distribution $f^{\rm NE}$ represents the statistics for
the electron population of an open quantum system, i.e. the central region (in the 
presence of interaction) connected to the two reservoirs. 
It is not the ``local'' equilibrium statistics of states scattering in and out of 
the reservoirs which are themselves at their own equilibrium.

From the definition $G^<=G^r \Sigma^< G^a$, where the total self-energy 
$\Sigma(\omega) = \Sigma_L(\omega) + \Sigma_R(\omega) + \Sigma_{\rm int}(\omega)$ 
arises from the contributions of the leads self-energy $\Sigma_{L,R}$ and 
the self-energy $\Sigma_{\rm int}$ of the interaction between particles, 
we can see that the total self-energy follows as well the same statistics, i.e.
$\Sigma^<  = -  f^{\rm NE}    \left( \Sigma^r - \Sigma^a  \right)$.
However as we have clearly explained in Ref.~[\onlinecite{Ness:2013}], there is no reason
for each contribution $\Sigma_{L,R}$ and $\Sigma_{\rm int}$ to follow individually the same statistics.

From this point of view, we find a compact and universal (with respect to the interaction)
expression for the NE distribution function $f^{\rm NE}(\omega)$:
\begin{equation}
\label{eq:fNE}
f^{\rm NE}(\omega) = \frac{f_0^{\rm NE}(\omega) - i \Sigma^<_{\rm int}(\omega)/\Gamma_{L+R}(\omega)}
{1+i (\Sigma^>_{\rm int}-\Sigma^<_{\rm int} ) / \Gamma_{L+R} } \ ,
\end{equation}
where $\Gamma_{L+R}(\omega)$ is the spectral function of the leads 
$\Gamma_{L+R}=\sum_{\alpha=L,R}  i(\Sigma_\alpha^r-\Sigma_\alpha^a)$, and 
$\Sigma^\lessgtr_{\rm int}$ are the lesser and greater components of the interaction self-energy.

The function $f_0^{\rm NE}(\omega)$ is the NE distribution for the non-interacting case. It can
be easily derived \cite{Schiller:1998,Ness:2009,Ness:2010,Ness:2012,Hershfield:1991} 
as the weighted average of the 
usual Fermi-Dirac distribution functions $f_{L,R}(\omega)$ of the left and right leads:
\begin{equation}
\label{eq:f0NE}
f_0^{\rm NE}
= \left(\Gamma_L(\omega)f_L(\omega) + \Gamma_R(\omega) f_R(\omega) \right) / \Gamma_{L+R}(\omega) \ .
\end{equation}
The distribution $f_0^{\rm NE}(\omega)$ is a double-step function, with more or less steep
steps (depending on the temperature) located around $\omega=\mu_L$ and $\omega=\mu_R$, and
separated by $\mu_L-\mu_R=eV$ ($\mu_\alpha$ being the chemical potential of the lead $\alpha=L,R$
and $V$ the applied bias).
The use of such a distribution has already been implemented in realistic calculations based on
single-particle elastic scattering \cite{Louis+Palacios:2003}.
 
The full NE distribution $f^{\rm NE}$ can be decomposed into two terms 
$f^{\rm NE}(\omega)=\tilde f_0^{\rm NE}(\omega)+\delta f^{\rm NE}(\omega)$,
one corresponds to the dynamically renormalized distribution 
$\tilde f_0^{\rm NE}= f_0^{\rm NE}(\omega) / \mathcal{N}(\omega)$
and the other is a ``correction'' term $\delta f^{\rm NE}$
associated with the inelastic processes and given by $\Sigma^<_{\rm int}$ renormalised by 
the same factor $\mathcal{N}$.
The renormalisation factor $\mathcal{N}(\omega)$ is given by the sum
$\mathcal{N}(\omega)=\Gamma_{L+R}+i(\Sigma^>_{\rm int}-\Sigma^<_{\rm int})$
of the spectral functions of the leads $\Gamma_{L+R}$
and of the interaction
$\Gamma_{\rm int}= i(\Sigma^>_{\rm int}-\Sigma^<_{\rm int})= i(\Sigma^r_{\rm int}-\Sigma^a_{\rm int})$,
and does not contain direct information of the statistics of the system.
 
For local electron-phonon interaction in the central region, the interaction self-energy 
is given by
$\Sigma_{\rm int}^{F,\lessgtr}(\omega) = \gamma_0^2 ( N_{\rm ph} G^\lessgtr(\omega \mp \omega_0) 
+ (N_{\rm ph} + 1) G^\lessgtr(\omega \pm \omega_0) )$ \cite{Dash:2010}. 
At low temperature $N_{\rm ph}=0$, and we can expand Eq.~(\ref{eq:fNE}) as a series expansion
in terms of the electron-phonon coupling parameter $\gamma_0$. 
To lowest order, we find the following expression for the NE distribution function:
\begin{equation}
\label{eq:fNE_LOE}
\begin{split}  
f^{\rm NE}(\omega) \sim &  f^{\rm NE}_0 + \frac{2\pi\gamma_0^2}{\Gamma}\ \times \\
 & [ A(\omega+\omega_0)\ (1-f^{\rm NE}_0(\omega))\ f^{\rm NE}_0(\omega+\omega_0) \\ 
 & - A(\omega-\omega_0)\ (1-f^{\rm NE}_0(\omega-\omega_0))\ f^{\rm NE}_0(\omega) ]
\end{split}  
\end{equation}
where $A(\omega)$ is the spectral function of the central region, i.e.
$A(\omega)=(G^a-G^r)/i2\pi$.
The terms in $\gamma_0^2$ in Eq.~(\ref{eq:fNE_LOE}) are correction terms to the non-interacting
distribution $f_0^{\rm NE}(\omega)$ and correspond to the lowest order contributions
of the electron-phonon interacting (i.e. phonon emission by electron or hole in the presence
of a finite bias). When they are included in the expression of the current \cite{Note:3} 
they generate an equivalent formulation of the lowest order treatment of the perturbation 
approaches to electron-phonon interaction provided in 
Refs.~[\onlinecite{Montgomery:2003a,Montgomery:2003b,Paulsson:2005}].

\section{Conclusion}
\label{sec:ccl}

We have demonstrated  that the NE density matrix developed by Hershfield for the steady 
state has the form of a McLennan-Zubarev non-equilibrium ensemble.
According to McLennan-Zubarev NESOM and Hershfield methods, the stationary density of an open 
system can be written in the modified Gibbs form $\rho^{\rm NE}=e^{-\beta(H-Y)}/Z$, 
with the non-equilibrium ``correction term'' $Y$.
The operator $Y$  that was interpreted as the operator into which $Y_0=\sum_i \mu_i N_i$ 
``evolves'' under the action of the perturbation $W$, is actually the entropy production rate of
the NE quantum system. It can be calculated in the absence and in the presence of interaction 
and gives information about the dissipation in the driven system.

The fact the both methods clearly show that a NE steady state can be mapped onto an
effective pseudo-equilibrium state, permits us to derive, in a rigorous way, NE quantities 
from the formal expressions given at equilibrium (compare Eq.~(\ref{eq:GF_feq}) and Eq.~(\ref{eq:GF_fNE}) ).
We have derived an example of such quantites, i.e. the NE distribution function for the electron population 
in a scattering region connected to two reservoirs.
Such a NE distribution function describes the statistics of an open quantum system in the NE steady
state regime. It is central to the understanding of the NE physical properties of open systems and 
to the derivation of NE thermodynamical laws, such as NE fluctuation-dissipation relations 
\cite{Ness:2013}, NE charge susceptibility \cite{Ness:2012} or quantum entropy production.


\begin{thebibliography}{55}
\bibitem{Note:3}
In the present set up, the electron current between the central region
and the left lead is given by the Meir and Wingreen formula. Using
Eq.~(\ref{eq:GF_fNE}), one introduces the NE distribution in the
definition of the current:
$I_L = (2ie/h)
\int {\rm d}\omega\ f_L \Gamma_L(\omega) [G^r(\omega)-G^a(\omega)] 
+ \Gamma_L(\omega)\ G^<(\omega)
\propto
\int {\rm d}\omega\ ( f_L - f^{\rm NE} )(\omega) \Gamma_L(\omega) A(\omega)$.
\end{thebibliography}
\end{document}